# A Rank Based Replacement Policy for Multimedia Server Cache Using Zipf-Like Law

Dr T R Gopalakrishnan Nair[1], P Jayarekha[2],

**Abstract**—The cache replacement algorithm plays an important role in the overall performance of Proxy-Server system. In this paper we have proposed VoD cache memory replacement algorithm for a multimedia server system. We propose a Rank based cache replacement policy to manage the cache space in individual proxy server cache. Proposed replacement strategy incorporates in a simple way the most important characteristics of the video and its accesses such as its size, access frequency, recentness of the last access and the cost incurred while transferring the requested video from the server to the proxy. We compare our algorithm with some popular cache replacement algorithm using simulation. The video objects are ranked based on the access trend by considering the factors such as size, frequency and cost. Many studies have demonstrated that Zipf's-like law can govern many features of the VoD and is used to describe the popularity of the video. In this paper, we have designed a model, which ranks the video on the basis of its popularity using the Zipf-like law. The video with higher ranking is named "hot", while the video with lower ranking is named "cold". The result show that the proposed rank based algorithm improves cache hit ratio, cache byte ratio and average request latencies compared to other algorithms. Our experimental results indicate that Rank based cache replacement algorithm outperforms LRU, LFU and Greedy Dual.

———————————— ◆ ————————————

## 1 Introduction

Recent advances in high speed networks and communication technologies have made it possible to provide an on-line access to a variety of information sources such as reference books, journals, newspapers images and video clips. A lot of performance degradations has been noticed due to exponential growth of on-line access traffic on the Internet. Various approaches are being explored by researchers to alleviate most of these problems. It is required to reduce user's perceived latency, improve server's performance and reduce traffic on the network. One attempt at reducing user's perceived latency is to move the source of information closer to the users, leading to the deployment of proxy servers which serve as intermediary between users and the central servers. User's request is forwarded to the central server by the proxy server, if it does not have a copy of the requested video in its cache. When the server returns the requested resource to the proxy server, the proxy, stores a copy in its cache while satisfying the user's request. Further requests to the same resource by the same user or another user are thus met at the proxy without contacting the central server again. This reduces the traffic on the network and the load on the servers. However it raises some performance issues:

1. If a request is to the video file already stored in the proxy cache space, what updation should be made to that video file's performance metric. [1]

2. When a request is to the video file which is not in the cache and if the cache storage space is fully consumed which of the video file currently in cache should be removed to give room for the incoming one.

Before we consider the above issues, we have to observe two important properties of the access pattern of video files.

*Property 1*

A small fraction of video files tend to be accessed much more often than the rest of the video files. This is shown by skewness of video file access probability distribution

In general, not every movie is equally accessed. Compared to other movies some of them are more frequently accessed, and therefore, are called hot

- *Dr.T.R.Gopalakrishnan Nair is the Director of Research and Industry Incubation Centre, DSI,Bangalore, India*
- *P Jayareka is Research Scholar, Dr. MGR University, Chenni. Working with the Dept. of Information Science and Engg, BMSCE, Bangalore, India.*

movies or most popular movies. In this paper we have used Zipf-like distribution, a general case of the zipf-law for the movie selection pattern.

*Property 2*

Sequential page access of video files. During normal playback, pages of the same video files are accessed in sequential order.

When a user wants to watch a video, the video is selected from an available list and within few seconds, the video is ready to be started. This provision of simple interface requires many complicated network structures which is usually made invisible to the users. A distributed VoD network [9] is designed using proxy servers in order to minimize latency at user-end and reduce the load at the server storing the movie library. It is useful to cache video in the proxy server in order to reduce startup latency for streaming service. For this reason, constructing an efficient proxy server cache is extremely important. There are many factors affecting the efficiency of a server cache, and the most critical one among these is the cache replacement policy.

## 2 Related work

A review of related studies is given in this section. At first, some traditional cache document replacement policies are discussed in the following.

Since the storage capacity of a cache proxy server is limited, it is impossible to hold all videos of the original central server in the proxy's cache. This rule is called the cache replacement policy. A predefined rule has to be defined to replace some cold videos. In order to maximize the hit rate it is advantageous to keep many small popular videos. However, to optimize the byte hit rate it is better to keep some large popular videos. There are two types of cache document replacement policies: single-factor cache document replacement policies and multiple-factor cache document replacement policies. Typical single-factor cache document replacement policies include the first-in-first-out (FIFO) policy, the random replacement (RAND) policy, the least recently used (LRU) policy, the least frequently used (LFU) policy, the LFU-aging policy,

- **LFU** - This policy keeps a reference count for each document in cache and when a document needs to be replaced, the document with the least reference count is removed. In a situation where there are two documents with the same reference count, it uses LRU to break the tie (i.e. it removes the least recently used among them).
- **Least Recently Used (LRU)**: Replaces the block in the cache that has not been used for the longest period of time. From the basics of temporal locality, the blocks that have been referenced in recent past will likely be referenced in the near future. This policy works well when there is a high temporal locality of references in the workload. An Early Eviction LRU (EELRU) proposed in [3], evicts the blocks when it notes that too many pages are being touched in a roughly cyclic pattern that is larger than the main memory.

- **Least Frequently Used (LFU)-Aging**: The LFU [5] policy can suffer from cache pollution (an effect of temporal locality): if a formerly popular object becomes unpopular, it will remain in the cache for a long time, preventing other newly or slightly less popular objects from replacing it. The aging policy is applied at intervals to bring down the reference counts of such objects and ultimately make them candidates for replacement. The LFU Aging policy is similar to the LFU policy. The least frequently used document is replaced when space is required for a new document. However, LFU-Aging attempts to deal with the problem of LFU. With LFU, some blocks can build up extremely high reference counts so that they are rarely (if ever) replaced, even if these blocks are never requested again. The LFU-Aging policy attempts to ensure that this condition does not occur by limiting and aging (i.e., occasionally reducing) reference counts. LFU-Aging addresses cache pollution when it considers both block access frequency and its age in cache. There is a variant called LFU with Dynamic Aging (LFUDA) that uses dynamic aging to accommodate shifts in the set of popular objects. It adds a cache age factor to the reference count when a new object is added to the cache or when an existing object is re referenced. LFUDA increments the cache ages when evicting blocks by setting it to the evicted object's key value. Thus, the cache age is always less than or equal to the minimum key value in the cache. In our approach a similar dynamic aging factor, not only helps in evicting a video which is currently offline, but also in balancing the frequency and time interval between the requests.

- **Random (RAND)**: It chooses among all blocks in the cache with equal probability. Intuitively, RAND seems appealing in this context if the client's caches are filtering all the locality characteristics from their reference streams. RAND provides a kind of lower bound on performance that is; there is no reason to use any policy that performs worse than RAND. Thus, this policy must be used if it is faster and less expensive.

Multiple-factor cache document replacement policies

- **Greedy Dual Size (GDS)**: It combines temporal locality, size, and other cost information. The algorithm assigns a cost/size value to each cache block [13]. In the simplest case the cost is set to 1 to maximize the hit ratio, but costs such as latency, network bandwidth can be explored. GDS assigns a key value to each object. The key is computed as the objects reference count plus the cost information divided by its size. The algorithm takes into account recency for a block by inflating the key value (cost/size value) for an accessed block by the least value of currently cached blocks. The GDS-aging version adds the cache age factor to the key factor. By adding the cache age factor, it limits the influence of previously popular documents.

- **The LRU-K policy**: The LRU-K policy is a variant of the LRU policy considering the last access time and access frequency of a document[18]. Two parameters, K and RP, have to be determined for the LRU-K policy. In the LRU-K policy, the last K access times of every document are recorded. When there is a necessity for cache document replacement, documents with access frequencies less than K will be replaced first. Subsequently, the LRU-K policy replaces documents that have not been requested recently, according to the LRU policy. If the time since the eviction of a document is greater than RP, then the access records of the document are deleted.

The organization of rest of the paper is as follows: In section 3. We present different Multimedia Server Architectures, Section 4 describes User behavior Pattern , In section 5 we present Proposed Approach and Algorithm, Section 6 presents the simulation results and discussion, Finally, in section 7, we conclude the paper and refer to further work.

## 3 Multimedia Server Architecture

Whenever a client requests for a video, a variety of metrics such as network load, content availability and distance are used to decide the most appropriate proxy server. The user's request is then redirected to that most suitable proxy server. The proxy server transmits the chosen content to the user via a dedicated unicast stream or through a multicast connection where many clients receive the same stream. Server push and client pull are the two methods of initiating delivery of video content from a proxy server to a client on a network. In client pull method it is necessary for the client to send a request in order to receive any video. Where as in server-push method, the server computers sends the video independent of requests from client computers. In this method a client makes an initial request to the server to start delivery of data. The server has complete control of how and when accesses to the storage media are executed. In this approach video is distributed to the proxy servers before any requests for this data have been made. If we anticipate that a specific vedio will be very popular (e.g., a blockbuster movie release), it is advantageous to distribute the object prior to any requests in order to avoid cache misses and reduce bandwidth consumption The server-push system is more appropriate for VoD system, while client-pull system is more suitable for accessing textual/numeric data[2]. Most conventional text file systems employ a client-pull architecture, in which the server retrieves information from disks only in response to an explicit read request from a client [1]. In server-push, the server continues to deliver data until the end of the stream is reached or the client requests some type of VCR operation.

Multimedia servers store and manage multimedia objects and deliver data streams in real-time, in response to requests from users. Additionally, they process the stored information before delivery to users. The content may range from long, sequentially accessed videos to composite documents consisting of a mixture of small multiple-media segments (e.g., video, audio, image, text). Multimedia servers may range from low-cost, PC-based simple servers that deliver a few streams, to scalable large (either centralized or distributed) servers that provide thousands of streams. Regardless of the server capacity, with given resources, the objectives of multimedia servers are to maximize the number of clients that can be simultaneously supported (high throughput), to minimize the latency between issuing request and initiation of display (low startup latency), and to support a smooth display without any artifacts, disruptions or jitters (continuous display).

Ideally, a server provides a standardized transparent interface to clients so that clients need not be aware of the specifics of the system (i.e., the hardware and software) that is providing the service. Today clients are often situated at workstations or on personnel computers, while servers are located elsewhere on the network, usually on more powerful machines.

There are three primary delivery architectures for the multimedia communication namely, peer-to-peer, central server and distributed servers.

The central server architecture (figure 1) has the ease of implementation for functionality, such as, security, privacy and billing. A majority of existing on-demand media streaming systems follow the Client-Server design, in which videos are stored in a set of central servers. All the requests are served by servers and the streaming content is directly delivered from servers. Obviously, this architecture is not scalable since the servers will become bottleneck as the requests increase. To save servers' resources and alleviate servers' traffic load, several proxy-based architectures have been proposed, in which a set of proxies are deployed in the network. Clients can request the cached portion of videos from the proxies. However, in both server-based solutions and proxy based solutions, the servers and proxies are expected to deliver high-quality streaming service to a large number of clients. Therefore, the servers and proxies should be very powerful in terms of computing power, upstream bandwidth, storage, etc., which makes the deployment and maintenance cost very expensive. On the other hand, recent research and experiments reveal that the current Internet has enough resources to support. The limitations of central server architecture are high delay, processing scalability and single point of failure [3]. And also centralized VoD systems have a limited streaming capacity.

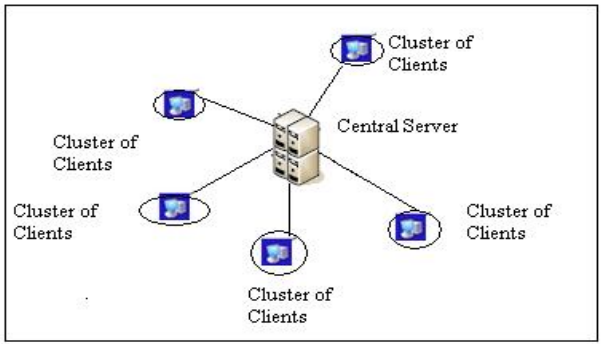

Fig. 1.Centralized Multimedia Server Architecture.

These shortcomings have prompted the development of distributed architectures with proxy servers (figure 2) installed at strategic locations in the network (closer to the clients). The proxy servers, located close to the user-end, cache content to reduce the load on the original server [4]. Proxy servers are usually placed between the clients and the content servers to reduce latency in repeated access and provide firewall protections. They have the disk cache space, network bandwidth, and availability to cache part of the objects for clients, making them good candidates to solve the bottleneck problem. However, large multimedia objects are not cached or only partially cached in current proxy servers, resulting in low cache efficiency. In a distributed Video-on-Demand system, customers are in charge of storing the video catalog, and they actively participate in serving video requests generated by other customers. The design of such systems is driven by key constraints like customer upload and storage capacities, video popularity distribution, and so on.

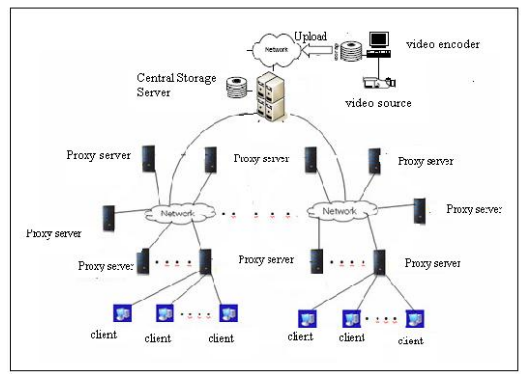

.  Fig 2  Distributed Multimedia server architecture

Peer-to-peer networks have gained a lot of popularity within the last two years (figure 3). Peers offer computing or storage capacity to others, while consuming from other peers at the same time. Consequently, a peer simultaneously acts as a client and a server [5]. The peer-to-peer architecture has low delay but is subject to high bandwidth usage. As a result, this architecture has serious access bandwidth problems, especially on upstream paths. However, their drawback is that users need to often wait for the full video to be downloaded before they can start watching it.

An extension of the proxy server approach is the use of content distribution networks (CDNs)(figure 4). In these networks, requests can be referred to other CDN servers (replicas or surrogates), which are generally located at the edge of the network core. CDNs can be seen as an extension of the proxy-based approach. Accordingly, the video requests are completely handled by edge servers, streaming the content directly to the clients. No requests are forwarded to the central server, as it instead happens in the proxy-based approach whenever the proxy does not have a copy of the requested content. Even if more robust than the centralized solution, major disadvantages limit the diffusion of the proxy-based and CDN approaches. The former *translates* a single point of failure into many points of failure, fractioning central server load to more servers. Additionally, all these systems have scalability problems; that is, when the number of clients increases, the only way to satisfy all the incoming requests is to add new servers proportionally.

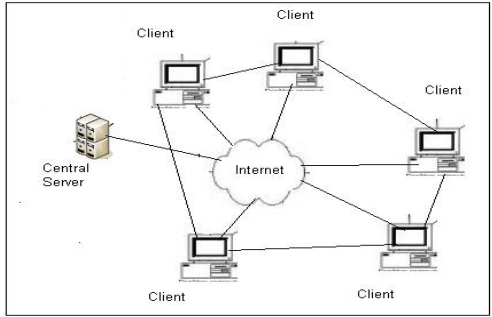

Fig. 3.. Peer-to-Peer Multimedia server architecture

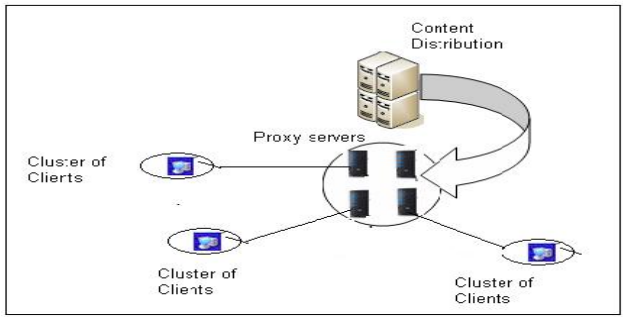

Fig 4  CDN Multimedia server architecture

## 4  User  Behavior Pattern

In order to acquire better system performance, users' behaviors must be well studied. For the video on demand delivery system, there are two most important user behaviors: the requests arrival pattern and the Video selection pattern. In this paper, we simulate the user behaviors as follows:

### 4.1 Request Arrival Pattern

The traditional Poisson distribution is generally used  to model the number of events occurring within  a given time interval. The formula for the Poisson probability mass function is $P(x,\lambda) = \dfrac{e^{-\lambda} \lambda^x}{x!}$ where $\lambda$ is the shape parameter that indicates the average number of events in the given time interval. But Hongliang Yu et al., [7] has proved that when the traditional Poisson model is used to represent the user-arrival pattern it is more conservative and overestimates the probability of large arrival groups. During some peak hours, the work load can be approximately five times greater than the normal one; if some special events occur, the rate can be even higher. Hence a modified  Poisson distribution model is introduced that more accurately models our observations. A modified Poisson distribution is obtained by replacing the independence variant x with (N − x), where N is the maximum number of user arrivals per unit of time. As Figure 5 shows, our tracking results match the modified Poisson distribution very well. This modified version of Poisson distribution can be  defined as:

$$P(x) = \frac{e^{-\lambda} \lambda^{(N-x)}}{(N-x)!}$$

X = 0, 1, 2... . We have assumed that workloads were generated with a range from 1 to 15 requests per minute in this paper. This range covers both light and heavy workloads and enables us to test performance under different traffic. Here, N is the maximum number of user arrivals in a target system. We used the values $\lambda$ = 15 and N = 27.

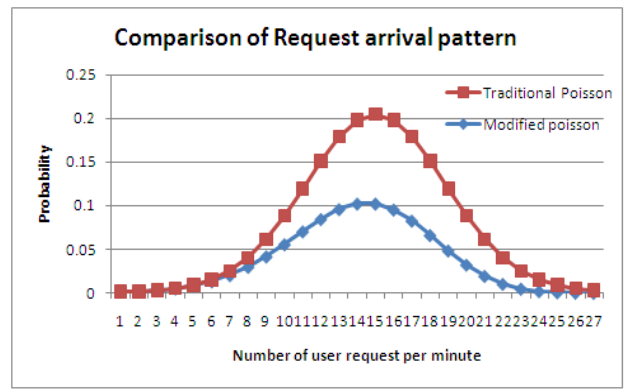

Fig 5 Request Arrival Pattern

### 4.2  Video selection pattern

In general, not every video is equally accessed. Some compared to other video some of them are more frequently accessed, and therefore, are called hot videos or most popular videos. In this paper we have used Zipf-like distribution, a general case of the Zipf-law for the movie selection pattern.

### Zipf law

Zipf's law is a famous statistical law that is observed in the behavior of many complex systems of different nature. It is a description of relationship between the frequency of occurrence of an event and its rank, when the events are ranked with respect to the frequency of occurrence. The law describes the phenomena where large events are rare, but small ones are quite common.

The law named for Harvard linguistic professor George Kingsley Zipf（1902-1950）was originally applied to the relationship between words in a text and their frequency of use. In 1994 Zipf's law was first employed to describe the web object access distribution in the work of Glassman.

Zipf's law states that the relative probability of a request for the $i^{th}$ most popular page is inversely proportional to $\frac{1}{i}$ , i.e.$n^{th}$ most popular video will be accessed with a probability equal to 1/n. The most frequently referenced resource is assigned a rank of 1, the Nth most frequently referenced resource is assigned a rank of N.

**Zipf-like law**

More studies done by Brealau gave the evidence that Web access pattern follows the Zipf-like law.

The distribution of page requests generally follow a Zipf-like distribution where the relative probability of a request for the ith most popular page is inversely proportional to $\frac{1}{i^{\alpha}}$ with $\alpha$ typically taking a value less than unity. Most of the reported estimates of $\alpha$ lie in the range between 0.6 and 1.0. Studies [12][13] have demonstrated that the value of parameter $\alpha$ is ranged between 0.75 and 0.85 for the servers, 0.64 and 0.83 for the proxies. In the discussion of this paper, we make use Zipf- like distribution.

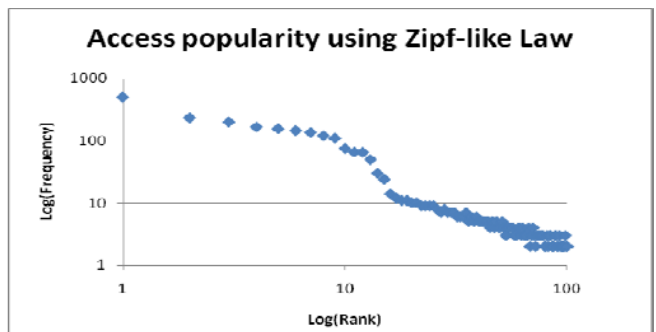

Fig 6 Zipf-Like Distribution

The above figure 6 shows that about only most popular 10% of videos are requested by 60% of the clients. Hence in our paper, these small numbers of popular videos are streamed from the proxy's cache. From the study [16] it has been suggested that a high proportion (70%) of sessions are terminated in the first 20 minute. Therefore in our paper the cache space is effectively utilized by dynamically allocating the cache buffer to the popular videos [17].

# 5 Proposed Approach and Algorithm

## 5.1 Overview of the proposed Approach

Compared with the cache objects in conventional web caching environment, multimedia caching has a different set of characteristics such as large volume, playback continuity, and being real-time. In addition, its cache replacement policy should consider the changes of object sizes incurred by users' reference patterns.

Frequency-based polices use object popularity (or frequency count) as the primary factor. From the literature survey, we can reach a conclusion that LFU replacement policy is the best one of all the tested replacement policies under the condition of multimedia object access characteristic satisfying the Zipf-like law. As a matter of fact, LFU replacement policy and Zipf-like law have the same basis, they all make use of frequency measurements, i.e. frequencies of multimedia video objects, the experiments indicate that LFU is one of the simple and effective replacement policies, which conform to many study results of Web caching research area [3][9]. A Rank value (RV) is assigned to each multimedia video object representing its importance in the cache. When the cache is full, the multimedia video object with the lowest RV is replaced. Our proposed approach tries to keep in the cache multimedia objects in the cache, those objects which take a long time to connect to server. The cost factor gives priority to those multimedia objects that must be loaded over the slowest Internet links, that have been referenced the most frequently, and that are small.

RV = Age + (cost /Size) *Pv

Age = $\left( \dfrac{t_{stay}}{t_{cur} - t_{ref}} \right)$ $t_{stay}$ is the staying time of video

in the cache buffer

$t_{cur}$ is the current reference time of video ,

$t_{ref}$ is the last reference time of video

Cost = $C_s + \dfrac{k}{B_s}$

$C_s$ =the time to connect with server $s$

$B_s$ = the bandwidth to server s
k is a constant
Size = Size of a block of video in bytes.

$$P_v = \frac{P_v^{1/\alpha}}{\left(Log_{10} size\right)^b}$$

$P_v$ = Probability of re-accessing the video block.

$P_v = V_{i+1}/V_i$

$V_i$ = the total number of videos seen so far which have been requested at least i times in the trace

$\alpha$ Characteristic value of Zipf's-like distribution set to 0.77

b A constant that weights the size factor

Fig 7    Expression for Rank Value (RV)

**5.2 Proposed Algorithm**

Algorithm RV Based Replacement

Initialize Age← 0.

Process each request to a video
The current request is for video p;
(1)  If p is already in cache memory M
(2)  RV(p) ←Age + (cost /Size) *Pv.
(3)  If p is not in the memory.
(4)     While there is not enough room in the memory for p
(5)     Let L ←$min_{q \in M}$ RV(q)
(6)     Evict q such that RV(q) = L
(7)  Bring p into memory and set
(8)  RV(p) =  Age + (cost /Size) * Pv
         End

The Age formula used in our algorithm takes a higher value whenever a video is referenced frequently and recently, a lower value when a video gradually becomes stale. This value has a direct impact on RV.

Our algorithm uses object size as one of the primary factor, and these usually remove larger video first. The rationale behind this is the fact that cache memory stores a small sized videos of most popular movies and removing a larger video can make room for multiple smaller ones. Therefore, it works well when large videos are less popular. This happens when users tend to access a small clip of popular movie than a long clip of less popular one. Hence size plays a representative factor in this algorithm.

It has been shown that web requests do not follow exactly Zipf's law. Brelau introduces a model for web requests, shown in equation 4, which follows a Zipf's-like behavior [6, 7].

Another major concern for proxy server is to reduce the latency of requests through caching, as numerous studies have shown that the waiting time has become the primary concern of VoD users. Our replacement algorithm called RV Based Replacement, which takes into account the different latencies incurred to load different prefixes, and attempts to minimize the average latency. Our algorithm enables to keep in the cache those videos from servers that take a long time to connect to, that must be loaded over the slowest Internet links. The simulation results show that in general our algorithm has a lower average latency than LRU, LFU Greedy Dual.

# 6  Results and Discussion

**6.1 Performance Metrics**

The performance metrics used to evaluate the various replacement policies used in this simulation are cache hit ratio and volume of byte transferred.

**Cache Hit Ratio**

This is the ratio of the number of requests met in the proxy's cache to the total number of requests. The higher this ratio the better the replacement policy being considered.

**Byte Volume Transferred**

This metric is used to characterize the amount of data flowing between the proxy and the servers. It is calculated as the ratio of total byte transferred between the proxy and the server to the total byte requested. The lower is this value, the better the replacement policy. It signifies a greater percentage of the requested byte that are met in the proxy's cache and they do not need to be loaded from the server [14]. Thus the best algorithm is s the one that maximizes cache hit ratio and minimizes byte volume transferred.

**Reduction in the Latency Time**.

Latency is defined as the delay between a request for a video and receiving the video. The latency problem occurs when users judge the download time as too long. Although an unacceptable latency adversely affects user satisfaction.

**6.2 Simulation Results**

This experiment is conducted to investigate the performance of the proposed replacement algorithm considering different varieties of videos. Different values of $\alpha$ was adopted for Zipf-like distribution. The parameter $\alpha$ determines the access pattern of the system. A larger value indicates that the accesses are biased to a certain set of popular videos. If the system has enough capacity to maintain these popular objects in the cache, it can achieve a high hit ratio. Consequently the byte-hit ratio increase with $\alpha$. When $\alpha$ exceeds 0.6, the popularity of a video becomes more concentrated RV Based Replacement algorithm caches all segments of the most popular videos, compared with (Cost Effective Replacement Algorithm) CERA [10] algorithm achieves higher byte rate.

Simulation results presented below are an average of several simulations conducted on the model

**Performance evaluation**

From our simulation results we have obtained the following analysis results

**Hit Ratio**. Our algorithm has proved that it can serve more number of clients compared with other traditional replacement algorithm. Our Experimental results as shown in figure 9 have achieved improvement of 10% in hit ratio.

**Byte Hit Ratio**. Our proposed algorithm to optimize the byte hit ratio by retaining some large popular videos. The Age formula used in our algorithm takes a higher value whenever a video is referenced frequently and recently, a lower value when a video gradually becomes stale. This value has a direct impact on RV.

Hence our algorithms has shown figure 8 an improvement of 8-10% in byte-hit ratio as compared with other algorithms.

**Reduction in latency time**

Our proposed algorithm is designed for a distributed VoD architecture using proxy servers in order to minimize latency at user-end and reduce the load at the Server as shown in figure 10. The latency time reduces as the cache size increases; our algorithm shows the best result as compared with other replacement policies.

**Comparison with CERA Algorithm**

Our Algorithm shows a better byte hit ratio as compared with CERA algorithm as shown in fig 7. As Compared with CERA our algorithm considers Age and the cost of transferring the larger multimedia video objects from server to the proxy.

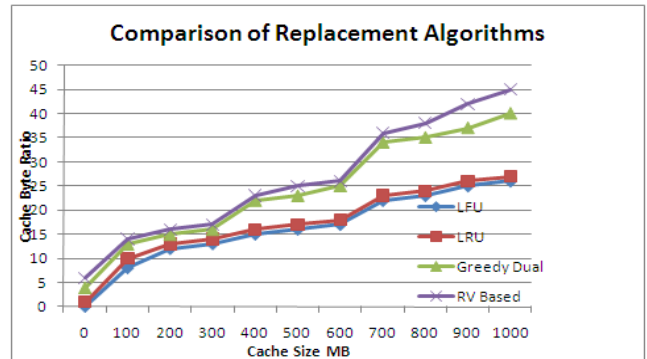

Fig 8 Comparison of Byte hit Ratio

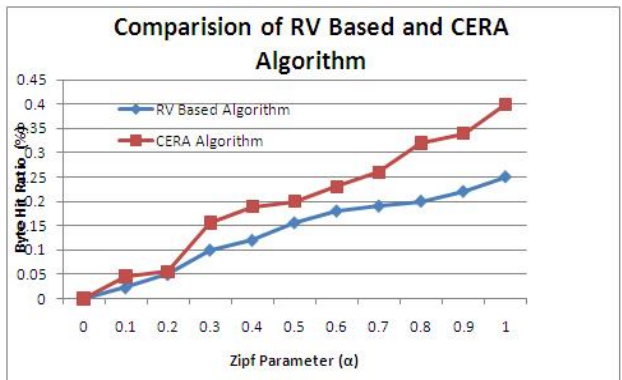

Fig 7. Comparison with CERA Algorithm

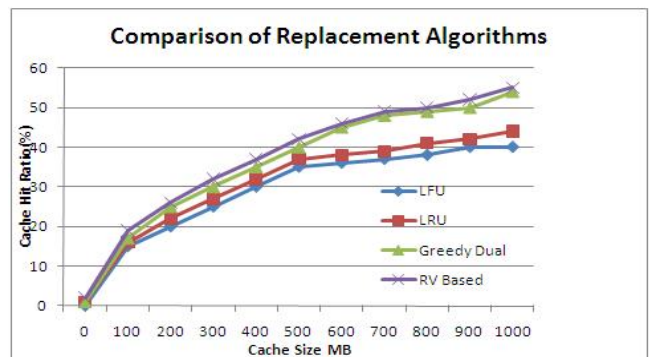

Fig 9 Comparison of Hit Rate

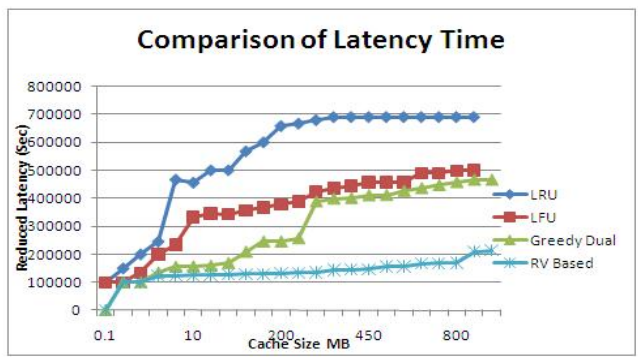

Fig 10 Reduction in latency time

# 7  Conclusion.

In this paper we have shown that by considering size, cost Age and Zipf-like law for selecting a victim for eviction, better results are obtained compared with other replacement algorithms. In our algorithm by balancing size and popularity of video object an increase in the byte hit ratio is obtained. Since cost factor uses the transferring time of video objects from server to proxy, our algorithm has shown good reduction in the latency time. Zipf-like law used to calculate Rank value retains most popular videos in the cache by which a significant improvement has achieved in hit ratio compared with traditional replacement algorithms.

Gopalakrishnan Nair holds M.Tech. (IISc, Bangalore) and Ph.D. degree in Computer Science. He has 3 decades experience in Computer Science and Engineering through research, industry and education. He has published several papers and holds patents in multi domains. He won the PARAM Award for technology innovation. Currently he is the Director of Research and Industry in Dayananda Sagar Institutions, Bangalore, India.

P Jayarekha holds M.Tech (VTU Belgaum ) in computer science securing second rank . She has one and a half  decades experience in teaching field. She has published many papers. Currently she is working as a teaching faculty in the department of Information science and engineering at BMS College Of Engineering , Bangalore ,India